\documentclass{icrc2009}

\usepackage{graphicx}   
\usepackage[caption=false]{caption}    
\usepackage[font=footnotesize]{subfig} 
\usepackage{fixltx2e}
\usepackage{url}

\newcommand{\shorttitle}[1]%
{\markboth{Proceedings of the 31\MakeLowercase{$^{st}$} ICRC, {\L}\'{o}d\'{z} 2009}{#1} }
\newcommand{\etal}{\MakeLowercase{\textit{et al. }}} 

\begin{document}
\title{High zenith angle observations of PKS\,2155-304 with the MAGIC telescope}

\author{\IEEEauthorblockN{D. Hadasch\IEEEauthorrefmark{1},
			  T. Bretz\IEEEauthorrefmark{2} and
                          D. Mazin\IEEEauthorrefmark{3} for the MAGIC collaboration}
                            \\
\IEEEauthorblockA{\IEEEauthorrefmark{1}Grup de F\'{\i}sica de les Radiacions,
Universitat  Aut\`onoma de Barcelona,
E-08193 Bellaterra, Spain}
\IEEEauthorblockA{\IEEEauthorrefmark{2}Universit\"at W\"urzburg, Am Hubland, D-97074 W\"urzburg, Germany}
\IEEEauthorblockA{\IEEEauthorrefmark{3}IFAE, Edifici Cn. UAB, E-08193 Bellaterra (Barcelona), Spain}}

\shorttitle{D. Hadasch \etal High zenith angle observations of PKS\,2155-304}
\maketitle

\begin{abstract}
The high frequency peaked BL Lac \mbox{PKS\,2155-304} with a redshift  z=0.116 was
discovered 1997 in the VHE range by the University of Durham Mark 6 \mbox{$\gamma$-ray}
telescope in Australia with a flux corresponding to $\sim$0.2 times the Crab Nebula
flux \cite{1}. It was later observed and detected with high significance by the
Southern observatories \mbox{CANGAROO} and H.E.S.S. establishing this source as the
best studied Southern TeV blazar. Detection from the Northern hemisphere was
very difficult due to challenging observation conditions under large zenith
angles.  In July 2006, the H.E.S.S. collaboration reported an extraordinary
outburst of VHE $\gamma$-emission \cite{2}. During the outburst, the VHE $\gamma$-ray
emission was found to be variable on the time scales of minutes and at a mean
flux of $\sim$7 times the flux observed from the Crab Nebula \cite{2}. The MAGIC
collaboration operates a 17m imaging air Cherenkov Telescope at La Palma
(Northern Hemisphere). Follow up observations of the extraordinary outburst
have been triggered in a Target of Opportunity program by an alert from the
H.E.S.S. collaboration. The measured spectrum and light curve are presented.
  
\end{abstract}

 
\section{Introduction} 
The 17m diameter MAGIC telescope on the Canary Island of
La Palma is the world's largest single imaging atmospheric Cherenkov telescope.
One of the aims of the MAGIC collaboration is to carry out observations at
large zenith angles. Under these special conditions and with a good telescope
sensitivity a bigger effective area is given and sources from a large section
of the Southern sky can be observed with a threshold of a few hundred GeV
($\approx$\,100\,GeV-500\,GeV, zenith angle dependent).

Here we present the results of an analysis of the Crab Nebula data set taken under large zenith angles (60$^\circ$ to
66$^\circ$). In addition to the usual image parameters \cite{3} timing information of the
recorded signals was used for reconstruction of the shower origin and for the
background suppression.
 
Furthermore we present a reanalysis of a \mbox{PKS\,2155-304}
data set recorded at a zenith angle range between 59$^\circ$ and 64$^\circ$. The same analysis
and cuts are used as for the Crab Nebula data set.
 
\section{Analysis}
All the data analysed in this work was taken in  wobble mode, i.e. tracking a
sky direction, which is 0.4$^\circ$ off the source position. The background
is estimated from the same field-of-view, which improves the background
estimation and yields a better time coverage because no extra OFF data have to
be taken.  

Compared to the previous analysis \cite{4} improvements are obtained
because of an updated Monte Carlo (MC) sample at high zenith angles leading to
a better \mbox{data-MC} agreement. A further improvement is achieved in the image
cleaning and in the gamma/hadron separation thanks to the usage of the timing information
of the images.  

In the new analysis we use the Time Image Cleaning \cite{5}: With a sub-nsec timing
resolution of the data acquisition system and thanks to the parabolic structure
of the telescope mirror a smaller integration window can be chosen. This
reduces the number of pixels with signals due to night sky background, which
survive the image cleaning.
This allows reducing the pixel threshold level (i.e. recorded charge) of the
image cleaning leading to a lower analysis energy threshold.  

For the analysis a robust set of dynamical cuts with a small number of free parameters is used
\cite{6}. In addition, cuts in time parameters are applied, which describe the time
evolution along the major image axis and the RMS of the time spread. These two
additional parameters lead to a better background suppression yielding a better
sensitivity of the analysis. The energy estimation is done with the Random
Forest regression method \cite{7}.

\section{Results}
\subsection{Crab Nebula}
The Crab Nebula is one of the best studied celestial objects because of the
strong persistent emission of the Nebula over 21 decades of frequencies. It was
the first object that was detected at TeV energies by the Whipple collaboration
\cite{8} in the year 1989 and is the strongest steady source of VHE $\gamma$-rays. Due to the
stability and the strength of the $\gamma$-ray emission the Crab Nebula is
generally considered the standard candle of the TeV \mbox{$\gamma$-ray}
astronomy. The
measured $\gamma$-ray spectrum extends from 60\,GeV \cite{9} up to 80\,TeV
\cite{10} and appears to be constant
over the years (from 1990 to present).

\subsubsection{Observation and Detection}
\begin{figure}[ht]
  \begin{center}
    \includegraphics*[width=0.49\textwidth,angle=0,clip]{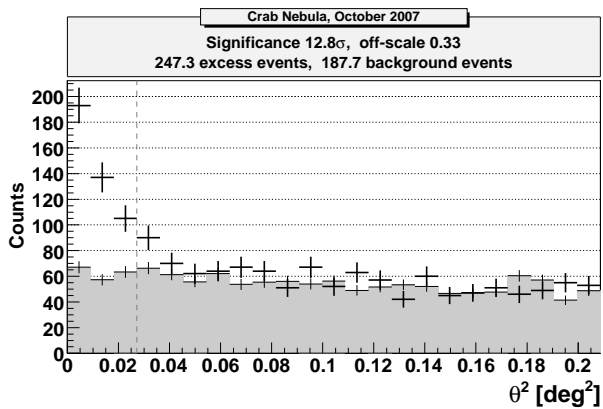}
    \caption{\label{fig:crab_theta}The On-Source and normalized background distribution of $\Theta^{2}$. The On-Source is shown in the black crosses and the background is shown in the gray shaded region. 2.22\,hrs of Crab data show an excess with a significance of 12.8$\sigma$.}
  \end{center}
\end{figure}

In October 2007, the MAGIC telescope took data of the Crab Nebula with a zenith
angle range of 60$^\circ$ up to 66$^\circ$. The data was taken under dark sky conditions and in
wobble mode. After quality cuts an effective on-time of 2.15\,hrs is obtained.
Using detection cuts (i.e. optimized on significance of a different Crab Nebula
sample), a total of 247 excess events above 187 background events with a
scale factor of 0.33 have been detected (see Fig.~\ref{fig:crab_theta}). According to Li\&Ma formula 17
\cite{11} the significance of this \mbox{$\gamma$-ray} signal is 12.8\,$\sigma$. This corresponds to an
analysis sensitivity of 8.7 $\frac{\sigma}{\sqrt{h}}$. Using the same set of cuts we obtain the
following sensitivities for integral fluxes ($\Phi$):
\begin{eqnarray*}
\Phi(E>0.4\,\mathrm{TeV}) &\Rightarrow& 5.7\% \mathrm{\,\,Crab\,\,in\,\,50\,hrs} \\
\Phi(E>0.63\,\mathrm{TeV}) &\Rightarrow& 5.6\% \mathrm{\,\,Crab\,\,in\,\,50\,hrs} \\
\Phi(E>1.0\,\mathrm{TeV}) &\Rightarrow& 5.9\% \mathrm{\,\,Crab\,\,in\,\,50\,hrs} \\
\Phi(E>1.5\,\mathrm{TeV}) &\Rightarrow& 6.8\% \mathrm{\,\,Crab\,\,in\,\,50\,hrs}
\end{eqnarray*}

\subsubsection{Differential Energy Spectrum}
\begin{figure}[ht]
  \begin{center}
    \includegraphics*[width=0.49\textwidth,angle=0,clip]{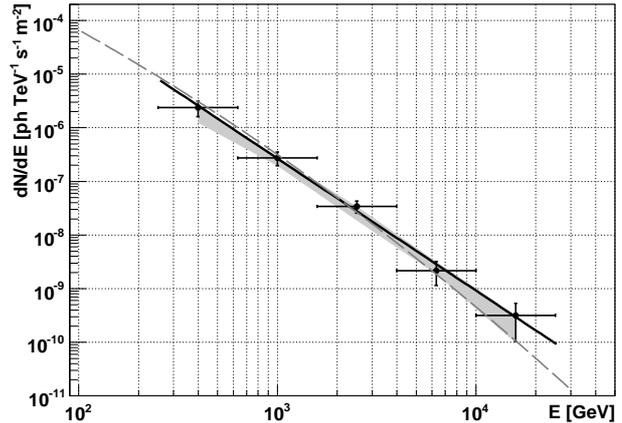}
    \caption{\label{fig:crab_spec}Differential energy spectrum of the Crab Nebula. Black line: power law fit to the data, gray band: systematic uncertanties, dashed line: published data taken at low zenith angles \cite{9}.}
  \end{center}
\end{figure}

The differential energy spectrum can be described well by a power law:
\begin{small}
\begin{eqnarray}
\frac{\mathrm{d}N}{\mathrm{d}E}=(2.7\pm0.4)\cdot10^{-7}\left(\frac{\mathrm{E}}{\mathrm{TeV}}\right)^{-2.46\pm0.13}\left(\frac{\mathrm{phe}}{\mathrm{TeV}\,\mathrm{s}\,\mathrm{m}^{2}}\right). \nonumber
\end{eqnarray}
\end{small}
The spectrum is shown in Fig.~\ref{fig:crab_spec}. The gray band represents the
range of results obtained by varying the total cut efficiency between 40\% and
70\%. For comparison, the Crab Nebula spectrum from
data taken at low zenith angles is drawn as a dashed line \cite{9}. A very good
agreement has been found.  

\subsection{PKS\,2155-304}
Like the Crab Nebula being the so-called standard candle of $\gamma$-ray astronomy,
the blazar PKS\,2155-304 is the so-called lighthouse of the Southern hemisphere. The
high frequency peaked BL Lac PKS\,2155-304 at a redshift of z=0.116 was
discovered in the VHE \mbox{$\gamma$-ray} range by the University of Durham Mark 6 $\gamma$-ray
telescope (Australia) in 1997 with a flux corresponding to $\sim$0.2 times the Crab Nebula
flux \cite{1}. It was later observed and detected with high significance by the
Southern observatories CANGAROO \cite{12} and H.E.S.S. \cite{13} establishing this source
as be the best studied Southern TeV blazar. Detection from the Northern
hemisphere is difficult due to challenging observation conditions under
large zenith angles. In July 2006, the H.E.S.S. collaboration reported an
extraordinary outburst of VHE $\gamma$-emission \cite{2}. During this outburst, the
$\gamma$-ray emission was found to be variable on time scales of minutes with a mean
flux of $\sim$7 times the flux observed from the Crab Nebula. Follow up observations
of the outburst by the MAGIC telescope have been triggered in a Target of
Opportunity program by an alert from the H.E.S.S. collaboration \cite{4}.

\subsubsection{Observation and Detection}
\begin{figure}[ht]
  \begin{center}
    \includegraphics*[width=0.49\textwidth,angle=0,clip]{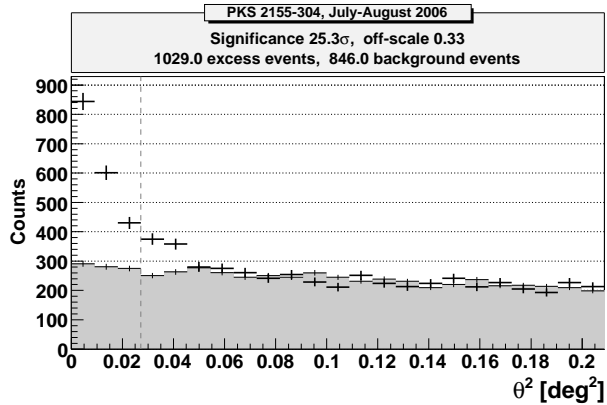}
    \caption{\label{fig:2155_theta}The On-Source and normalized background distribution of $\Theta^{2}$. The denotations are the same as in figure \ref{fig:crab_theta}. A clear excess with a significance of more than 25 standard deviations for a source at the position of PKS\,2155-304 is found.}
  \end{center}
\end{figure}

The MAGIC telescope observed the blazar \mbox{PKS\,2155-304} from 28 July to 2 August
2006 in a zenith angle range from 59$^\circ$ to 64$^\circ$. The data were taken under dark sky
conditions and in wobble mode. After quality cuts a total effective on-time of
8.7\,hrs is obtained. For the detection of \mbox{PKS\,2155-304}, the same cuts are used as
for the detection of the Crab Nebula. Three OFF regions are used and 1029
excess events above 846 background events are detected. A significance of 25.3
standard deviations is obtained. The corresponding $\Theta^{2}$-plot is presented in Fig.~\ref{fig:2155_theta}.

\subsubsection{Differential Energy Spectrum}
The differential energy spectrum is shown in Fig. 4 as a black line together
with the measured spectrum of H.E.S.S. during the strong
outburst \cite{2} (dashed line). Note that H.E.S.S. and MAGIC data are not simultaneous. The
obtained spectral points in this analysis are fitted from 400\,GeV on, because at
lower energies H.E.S.S. reported a change of the slope ($-3.53\pm0.05$ above
400\,GeV to $-2.7\pm0.06$ below 400\,GeV). The fitted MAGIC data points are consistent
with a power law:
\begin{small}
\begin{eqnarray}
\frac{\mathrm{d}N}{\mathrm{d}E}=(1.8\pm0.2)\cdot10^{-7}\left(\frac{\mathrm{E}}{\mathrm{TeV}}\right)^{-3.5\pm0.2}\left(\frac{\mathrm{phe}}{\mathrm{TeV}\,\mathrm{s}\,\mathrm{m}^{2}}\right) \nonumber
\end{eqnarray}
\end{small}
with a fit probability after the $\chi^{2}$-test of 81\%.  Above 400\,GeV, the
energy spectrum measured by H.E.S.S. from the preceding flare of PKS\,2155-304
is one order of magnitude higher than the spectrum measured by MAGIC, but the
spectral slope ($-3.53\pm0.05$) is consistent within the
statistical errors.
\begin{figure}[ht]
  \begin{center}
    \includegraphics*[width=0.49\textwidth,angle=0,clip]{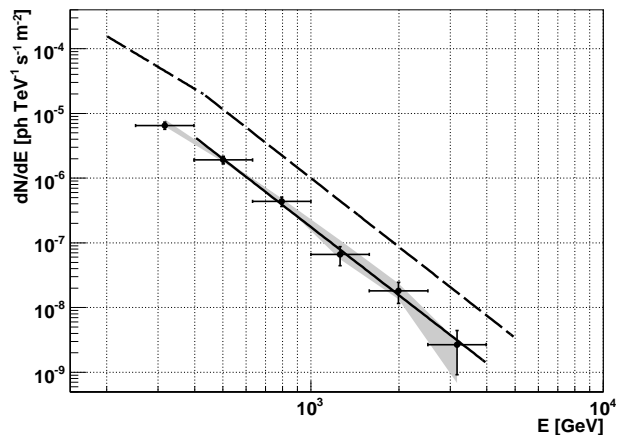}
    \caption{Differential energy spectrum (black line) together with
systematic errors due to varying cuts efficiencies (gray band). The black
dashed line corresponds to the H.E.S.S. measurement during the flare.}
  \end{center}
\end{figure}
\begin{figure}[ht]
  \begin{center}
    \includegraphics*[width=0.49\textwidth,angle=0,clip]{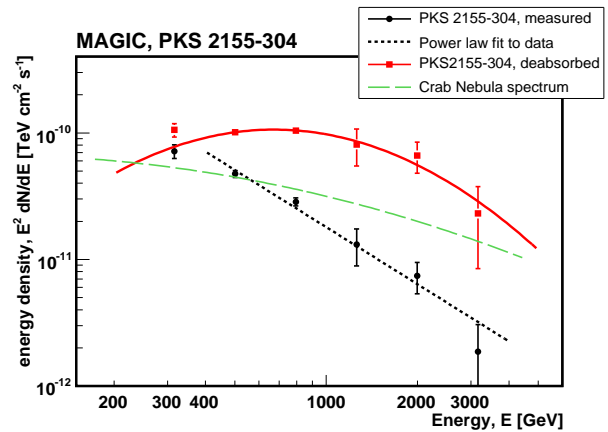}
\caption{\label{fig:2155_deab}Measured and intrinsic differential energy density of PKS\,2155-304. The effect of the EBL is taken into account by using the recent model of Kneiske adopted by MAGIC \cite{3c279}. The power law fit to the observed spectrum is given by the dotted line, while the curved fit to the intrinsic spectrum is shown by the solid line. The fitted peak position is located at $E_{peak}=(672^{+104}_{-157})\,\mathrm{GeV}$. For comparison the Crab Nebula spectrum is shown as the dashed line.}
  \end{center}
\end{figure}
\subsubsection{Intrinsic Energy Spectrum}
The VHE photons of PKS\,2155-304 interact with the low-energy photons of the
extragalactic background light (\cite{gould, hauser}). The predominant reaction
$\gamma_{VHE}+\gamma_{EBL}\rightarrow e^{+}e^{-}$ leads to an attenuation of
the intrinsic spectrum $\mathrm{d}N/\mathrm{d}E_{intr}$ that can be
described by
\begin{small}
\begin{eqnarray}
\mathrm{d}N/\mathrm{d}E_{obs}=\mathrm{d}N/\mathrm{d}E_{intr} \cdot \exp[-\tau_{\gamma \gamma}(E, z)] \nonumber
\end{eqnarray}
\end{small}
with the observed spectrum $\mathrm{d}N/\mathrm{d}E_{obs}$, and the energy
dependent optical depth $\tau_{\gamma \gamma}(E, z)$. Here we use the recent
 model of Kneiske et al. that has been adopted by MAGIC \cite{kneiske,3c279}. The measured spectrum and
the reconstructed deabsorbed spectrum are shown in Fig.~\ref{fig:2155_deab}.
For comparison the Crab Nebula spectrum is also shown. 
A power law fit to the deabsorbed spectrum results in a spectral index of
$2.4\pm0.1$. However, the fit probability is rather low (4\%) which motivates a
higher order fit function. We have chosen a curved power-law fit (a parabolic shape
in log-log representation) of the following form:
$\mathrm{d}N/\mathrm{d}E=N_{0}(E/1\mathrm{TeV})^{-\alpha + \beta \cdot
\ln(E/1\mathrm{TeV})}$.  The best fit parameters are: \mbox{$N_{0}=(9.7 \pm
0.9)\cdot10^{-11} \,\mathrm{TeV}^{-1}\mathrm{cm}^{-2}\mathrm{s}^{-1}$,}
\mbox{$\alpha=2.4\pm0.2$}, $\beta=-0.5\pm0.2$ (solid line in Fig.~\ref{fig:2155_deab}) and the fit probability is 77\%. The observed curvature is indicative of
a maximum in the energy density and is usually interpreted as due to inverse
Compton (IC) scattering.  The fitted peak position is determined to be at
$E_{peak}=(672^{+104}_{-157})\,\mathrm{GeV}$.

\subsubsection{Light curve}
\begin{figure*}[ht]
  \begin{center}
    \includegraphics*[width=0.9\textwidth,angle=0,clip]{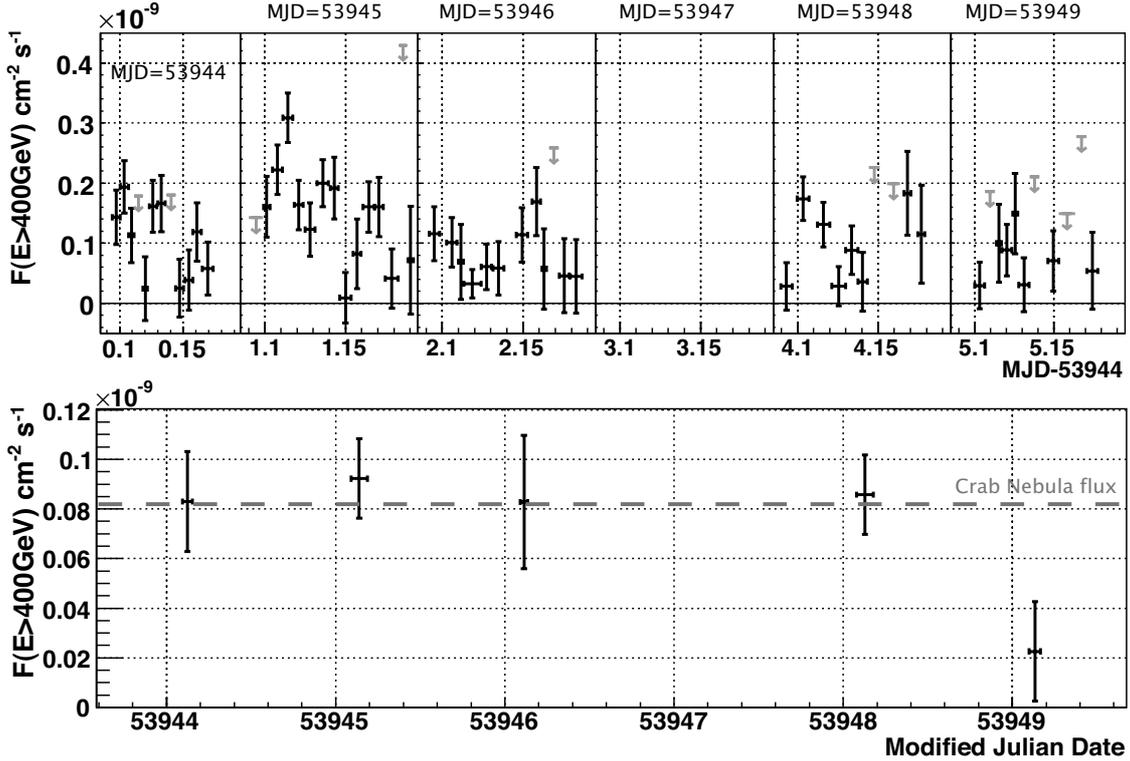}
    \caption{\label{fig:lc}\textit{Top:} Light curve for individual nights for E$>$400\,GeV of PKS\,2155-304. Only the second
night (MJD\,53945) shows significant variability. Vertical arrows represent flux
upper limits. \textit{Bottom:} Light curve for the total data set for E$>$400\,GeV with one
flux point per night. The error bars in x-direction represent the observation
time. The Crab Nebula flux is shown for comparison. For more information see text.}
 \end{center}
\end{figure*}

The integral light curves above 400\,GeV shown in Fig. \ref{fig:lc}
have a binning of one flux point per night (bottom
panel) and a binning in two runs which corresponds to about 10\,minutes per
bin (top panel).  
Significant detections in most of the time bins are obtained.
A significant intranight variability is found for the second night MJD\,53945 (29
July 2006) giving a probability for a constant flux of less than $5\cdot10^{-9}$. For the
other nights, no significant intranight variability is found. In the lower panel of Fig. \ref{fig:lc},
a night-by-night light curve is shown. 
A fit by a constant to the run-by-run light curve results in a chance probability of less than $10^{-12}$. 
However, a fit by a constant to the night-by-night light curve results in a chance probability of $7\times10^{-2}$.
We, therefore, conclude that there is a significant variability on the time scales reaching from days (largest scale we probed)
down to 20 minutes (shortest scale we probed).

\section{Conclusion}
A study of high zenith angles (60$^\circ$ - 66$^\circ$) observations with the
MAGIC telescope was performed. A new Time Image Cleaning and also time
parameters were used for the background suppression, which leads to a significant improvement of the sensitivity.

From Crab Nebula observations a sensitivity of 5.7\% of the Crab Nebula flux
for 50\,hrs of observations above 0.4\,TeV has been determined.
The differential energy spectrum of the Crab Nebula is
in excellent agreement with the published data at lower zenith angles.
This improved analysis is used to reanalyze data of \mbox{PKS\,2155-304} taken with MAGIC in 2006.

The energy spectrum from 400\,GeV up to 4\,TeV has a spectral index of
($-3.5\pm0.2$) and shows no change of spectral slope with flux state. It agrees
well with the results of  H.E.S.S. \cite{2} and CANGAROO \cite{12}.
Furthermore we corrected the measured spectrum for the effect of the EBL
absorption using the recent model of Kneiske et al. The resulting intrinsic spectrum
shows a clear curvature. A fitted peak position in the energy density
distribution is at $E_{peak}=(672^{+104}_{-157})\,\mathrm{GeV}$.

The light curves show a significant variability on daily as well as on intra-night time scales.
Finally we conclude that high zenith angle observations with the MAGIC telescope have proven to yield high quality spectra and light curve at a low energy threshold.

\section*{Acknowledgements}
We would like to thank the Instituto de Astrofisica de 
Canarias for the excellent working conditions at the 
Observatorio del Roque de los Muchachos in La Palma. 
The support of the German BMBF and MPG, the Italian INFN 
and Spanish MICINN is gratefully acknowledged. 
This work was also supported by ETH Research Grant 
TH 34/043, by the Polish MNiSzW Grant N N203 390834, 
and by the YIP of the Helmholtz Gemeinschaft.
D.M's research is
supported by a Marie Curie Intra European Fellowship within the 7th European
Community Framework Programme.

\end{document}